\documentclass[twocolumn,superscriptaddress,longbibliography,nofootinbib,aps,pre]{revtex4-2}

\usepackage{graphicx}
\usepackage{dcolumn}
\usepackage{bm}
\usepackage{amsmath,amssymb,mathtools}
\usepackage{hyperref}

\hypersetup{colorlinks=true,linkcolor=blue,citecolor=blue,urlcolor=blue}

\newcommand{\ee}{\mathrm{e}}
\newcommand{\ii}{\mathrm{i}}

\begin{document}

\title{Compact Dynamical Mean-Field Theory of Oscillator Networks}

\author{Kanishka Reddy}
\email{kani@uw.edu}
\affiliation{Department of Applied Mathematics, University of Washington}

\begin{abstract}
We present a compact dynamical mean-field theory (DMFT) for large networks of coupled phase oscillators whose phases live on the circle $S^1$ and interact with both coherent mean-field coupling and quenched randomness. Starting from wrapped Langevin dynamics, we build a path-integral representation that keeps the $2\pi$-periodicity of the phases explicit. After averaging over the disorder in the thermodynamic limit, this construction reduces to a single-oscillator stochastic equation driven by a deterministic mean field and a self-consistent colored Gaussian noise, whose covariance is fixed by a circular two-time correlator. In the limit of vanishing disorder, the formalism reproduces the Ott--Antonsen reduction and recovers standard Kuramoto and theta-neuron neural-mass equations. The same framework accommodates arbitrary $2\pi$-periodic coupling functions, including those obtained from infinitesimal phase response curves (iPRCs) of biophysical neuron models. As an example, we show that for adaptive exponential integrate-and-fire neurons, inserting an iPRC-fitted coupling into the compact DMFT yields quantitative predictions for synchronization thresholds, providing a direct route from single-neuron phase response data to network-level mean-field predictions for arbitrary phase-reducible oscillators.
\end{abstract}
\maketitle

\section{Introduction}
\label{sec:intro}

Networks of coupled phase oscillators are a standard tool for describing collective dynamics across physics, biology, and engineering. They capture synchronization phenomena in systems such as Josephson junction arrays, power grids, neural circuits, and circadian rhythms~\cite{Kuramoto1975,Strogatz2000,Acebron2005,Rodrigues2016,Wiesenfeld1998,BuzsakiWang2012}. In all-to-all ensembles with rank-one coupling, classical mean-field arguments already explain how macroscopic synchrony emerges, and the reductions due to Ott and Antonsen (OA) and to Watanabe and Strogatz (WS) make this picture precise: under suitable assumptions, the infinite-dimensional dynamics collapses onto a low-dimensional manifold parametrized by a complex order parameter~\cite{Ott2008,Watanabe1993,OttAntonsen2009}. Closely related constructions underlie neural-mass descriptions of theta-neuron and quadratic integrate-and-fire (QIF) populations, where the firing-rate dynamics of a large ensemble can be written as a pair of coupled ordinary differential equations~\cite{Montbrio2015,PazoMontbrio2014,Laing2014,BickReview2020}. These results together give a sharp and analytically tractable account of synchronization in idealized mean-field architectures.

Real networks, however, are seldom rank-one or uniform. Coupling matrices in neural circuits, power systems, and other applications are typically dense, heterogeneous, and often strongly nonreciprocal, shaped by quenched disorder that does not self-average in the thermodynamic limit. In such settings the effective input to a given oscillator is no longer a static mean field but a temporally correlated fluctuation whose statistics are self-generated by the network. Recent work on fully disordered Kuramoto-type ensembles has mapped out a rich phenomenology, including nontrivial incoherent states, volcano-type transitions of the local-field distribution, and collective chaos~\cite{OttinoLoefflerStrogatz2018PRL,PazoGallego2023PRE,KatiRanftLindner2024PRE,HongMartens2022Chaos,LeonPazo2022PRE,LeonPazo2025Chaos,BickPanaggioMartens2018,OlmiPolitiTorcini2010}. Analytical progress has often relied on numerical simulations or perturbative expansions in the coupling strength.

An important step toward a nonperturbative treatment was recently taken by Pr\"user, Rosmej, and Engel, who used dynamical-cavity methods to derive a self-consistent single-oscillator description of the fully disordered Kuramoto model~\cite{Prueser2024PRL,PrueserEngel2024PRE}. Their approach yields a single-site stochastic equation driven by colored noise, whose two-time correlator must be determined self-consistently, and provides quantitative predictions for the volcano transition in the local-field distribution.

In this paper we develop a more general framework, compact dynamical mean-field theory, that extends the dynamical-cavity construction of Pr\"user and Engel in three respects.

\emph{(i) Compact path-integral formulation on $S^1$.}
In Refs.~\cite{Prueser2024PRL,PrueserEngel2024PRE} the phase is treated as a noncompact variable on $\mathbb{R}$, and $2\pi$-periodicity is only enforced at the level of observables. Here we instead start from wrapped Langevin dynamics on the circle $S^1$ and build a Martin–Siggia–Rose–Janssen–de Dominicis (MSRJD) path integral in which the response field is integer-valued. This integer structure comes from the Fourier representation of the periodic Dirac comb enforcing identification modulo $2\pi$. Villain resummation~\cite{Villain1975,SiebererWachtelAltmanDiehl2016} turns the resulting integer sum into a periodic Gaussian, giving a continuum theory that keeps track of winding sectors and respects the topology of $S^1$ already at the level of the generating functional. In this compact formulation intrinsic phase noise $D$ and frequency disorder $\Delta$ are treated on the same footing, unlike in the dynamical-cavity approach.

\emph{(ii) Unified mean and random channels with arbitrary harmonics.}
Pr\"user and Engel work with purely random coupling and a fixed sinusoidal interaction. By contrast, we split the coupling matrix into a coherent mean $J_0/N$ and a zero-mean Gaussian random part of variance $g^2/N$. This gives a continuous interpolation between classical mean-field synchronization and disorder-dominated dynamics. The compact framework also allows general $2\pi$-periodic coupling functions $H(\Delta\theta) = \sum_m h_m e^{\ii m\Delta\theta}$, rather than only the $\sin(\Delta\theta)$ term of the Kuramoto model. In the limit $g\to 0$ the random channel switches off and the theory reduces exactly to Ott--Antonsen dynamics, recovering both Kuramoto ensembles and theta-neuron neural-mass models~\cite{Montbrio2015,Luke2013}.

\emph{(iii) Biophysical parameterization via phase response curves.}
The multiharmonic structure in (ii) has a natural biophysical interpretation. For any neuron model with a stable limit cycle, classical phase-reduction theory~\cite{BrownMoehlisHolmes2004} defines an infinitesimal phase response curve (iPRC) that describes how weak perturbations advance or delay the phase. The Fourier coefficients of the iPRC then directly determine the coupling function $H(\Delta\theta)$ in our description. This gives a concrete route from single-cell measurements to network-level DMFT parameters. As an illustration, we consider adaptive exponential integrate-and-fire (AdEx) neurons and show that Ott--Antonsen dynamics with iPRC-derived coupling reproduces synchronization thresholds in networks of phase-reduced AdEx oscillators.

Taken together, points (i)–(iii) place compact DMFT at the interface between the theory of disordered oscillator networks and biophysically motivated neural modeling. In contrast to voltage-based mean-field approaches for conductance-driven networks~\cite{Augustin2017,Zerlaut2017,diVolo2019,Carlu2020}, which start from a Fokker--Planck equation in voltage space and then derive rate equations, we first perform phase reduction, extract the coupling coefficients from the iPRC, and only then close the theory via DMFT. This sacrifices detailed subthreshold voltage statistics but yields a relatively simple analytical structure in which synchronization thresholds and related quantities follow from a small set of Fourier coefficients.

The remainder of the paper is organized as follows.
Section~\ref{sec:model} introduces the oscillator model and the coupling decomposition.
Sections~\ref{sec:compact_msrjd}--\ref{sec:disorder_avg} develop the compact path-integral formulation, perform the disorder average, and derive the effective single-site SDE.
Section~\ref{sec:oa_limit} shows how the Ott--Antonsen reduction and theta-neuron/QIF dynamics appear in the mean-channel limit.
Section~\ref{sec:iprc} details the iPRC-based parameterization using AdEx neurons.
Section~\ref{sec:numerics} compares the DMFT predictions with direct network simulations.
Section~\ref{sec:conclusion} summarizes the main results and outlines possible extensions.

\section{Model}
\label{sec:model}

We consider a network of $N$ phase oscillators $\theta_i(t)\in[-\pi,\pi)$, $i=1,\dots,N$, evolving according to the stochastic differential equation
\begin{equation}
\label{eq:model}
\dot{\theta}_i(t) = F_i(\theta(t),t) + \sqrt{2D}\,\xi_i(t),
\end{equation}
where $\xi_i(t)$ are independent white noises with
\begin{equation}
\langle \xi_i(t)\xi_{i'}(t')\rangle = \delta_{ii'}\delta(t-t'),
\end{equation}
and $D\ge 0$ controls the noise strength. The deterministic drift is
\begin{equation}
\label{eq:drift}
F_i(\theta(t),t) := \omega_i + I_i(t) + \sum_{j=1}^N W_{ij}\,H\!\big(\theta_i(t)-\theta_j(t)\big),
\end{equation}
where $\omega_i$ is the intrinsic frequency of oscillator $i$, $I_i(t)$ is an external drive, and $H(\Delta)$ is a $2\pi$-periodic interaction function.

We expand $H$ in a Fourier series,
\begin{equation}
\label{eq:harmonic}
H(\Delta) = \sum_{m\ge 1}\Big(h_m\,\ee^{\ii m\Delta} + h_m^{*}\,\ee^{-\ii m\Delta}\Big), \qquad h_m\in\mathbb{C}.
\end{equation}
The classical Kuramoto model is obtained by taking $h_1 = -\ii/2$ and $h_{m>1}=0$, which gives $H(\Delta)=\sin\Delta$. Keeping higher harmonics $m\ge 2$ allows for more general interactions, including those that arise from phase reduction of biophysical neuron models~\cite{BrownMoehlisHolmes2004,Daido1996,Skardal2011}. In Sec.~\ref{sec:iprc} we explain how the coefficients $h_m$ can be computed directly from an empirically measured or numerically obtained infinitesimal phase response curve.

The coupling matrix $W_{ij}$ is split into a uniform mean part and a quenched random part with the usual thermodynamic scaling:
\begin{equation}
\label{eq:coupling_split}
W_{ij} = \frac{J_0}{N} + g\,\widetilde{W}_{ij},
\end{equation}
where
\begin{equation}
\label{eq:random_stats}
\langle \widetilde{W}_{ij}\rangle = 0, \qquad 
\langle \widetilde{W}_{ij}\widetilde{W}_{i'j'}\rangle = \frac{1}{N}\,\delta_{ii'}\delta_{jj'}.
\end{equation}
Here $J_0$ sets the strength of the coherent mean-field coupling, and $g$ tunes the amplitude of quenched disorder. The variance structure $\delta_{ii'}\delta_{jj'}$ means that $\widetilde{W}_{ij}$ and $\widetilde{W}_{ji}$ are statistically independent, i.e. the random part is fully asymmetric. The $1/N$ scaling ensures that the cumulative effect of fluctuations remains $O(1)$ as $N\to\infty$. Varying $(J_0,g)$ interpolates between the classical Kuramoto case ($g=0$, $J_0>0$) and fully disordered ensembles ($J_0=0$, $g>0$).

The harmonic expansion~\eqref{eq:harmonic} naturally introduces collective order parameters
\begin{equation}
\label{eq:order_params}
Z_m(t) = \frac{1}{N}\sum_{j=1}^N \ee^{\ii m\theta_j(t)}, \qquad m\ge 1,
\end{equation}
which measure $m$-fold clustering. In particular, $|Z_1|$ is the usual Kuramoto order parameter: $|Z_1|=0$ corresponds to an incoherent state, while $|Z_1|>0$ indicates macroscopic synchronization.

For the intrinsic frequencies $\omega_i$ we will often assume a Lorentzian distribution,
\begin{equation}
\label{eq:lorentzian}
p(\omega) = \frac{\Delta}{\pi\big[(\omega-\omega_0)^2 + \Delta^2\big]},
\end{equation}
with center $\omega_0$ and half-width $\Delta>0$. This choice keeps many calculations simple (integrals over $\omega$ can be done by residues), but the DMFT construction itself does not rely on a single specific form of $p(\omega)$.

Because the phases live on the circle $S^1$, some care is needed when setting up a path-integral description. Standard MSRJD field theory uses a real-valued response field and implicitly assumes a noncompact target space. In the next section we construct a \emph{compact} MSRJD formulation that respects $2\pi$-periodicity and keeps winding sectors explicit in the continuum limit.

\section{Compact MSRJD Path Integral and Villain Resummation}
\label{sec:compact_msrjd}

We now construct a path-integral representation of the dynamics~\eqref{eq:model} that keeps the compact geometry of $S^1$ explicit. Phases are only defined modulo $2\pi$, so the usual MSRJD formalism with a real response field conjugate to an unconstrained coordinate has to be adapted.

\subsection{Discrete-time wrapped dynamics}
\label{sec:discrete_wrapped}

To fix ideas, start with a single phase $\theta_t\in[-\pi,\pi)$ updated on a discrete time grid $t\in\{t_0, t_0+\varepsilon, \dots\}$ with an It\^o scheme,
\begin{equation}
\label{eq:discrete_update}
\theta_{t+\varepsilon} = \theta_t + \varepsilon\,F(\theta_t,t) + \sqrt{2D\varepsilon}\,\eta_t + 2\pi\ell_t, \quad \ell_t\in\mathbb{Z},
\end{equation}
where $\eta_t\sim\mathcal{N}(0,1)$ are independent Gaussians. The integer $\ell_t$ enforces the wrap: after each update the phase is returned to $[-\pi,\pi)$ by adding the appropriate multiple of $2\pi$.

This wrapping is encoded by the periodic Dirac comb
\begin{equation}
\label{eq:dirac_comb}
\delta_{2\pi}(x) = \sum_{p\in\mathbb{Z}}\delta(x - 2\pi p)
                 = \frac{1}{2\pi}\sum_{k\in\mathbb{Z}}\ee^{\ii k x},
\end{equation}
where the second equality is the Poisson-summation (Fourier series) representation. Inserting~\eqref{eq:dirac_comb} into the path measure and integrating out the Gaussian noise $\eta_t$ yields, for each time slice, a weight that involves a sum over an integer $k_t$.

\subsection{Integer-valued response field}
\label{sec:integer_response}

Introducing sources $J_t$ for observables $O(\theta_t)$ and performing the Gaussian average over noise, we obtain the compact MSRJD generating functional
\begin{align}
\label{eq:compact_msrjd}
\mathcal{Z}[J] &\propto \int\!\mathcal{D}\theta \sum_{\{k_t\in\mathbb{Z}\}} \exp\bigg\{\sum_t \Big[\ii k_t\big(\Delta_\varepsilon\theta_t - \varepsilon F(\theta_t,t)\big) \nonumber\\
&\qquad\qquad\qquad\qquad - D\varepsilon\,k_t^2\Big] + \varepsilon\sum_t J_t O(\theta_t)\bigg\},
\end{align}
where $\Delta_\varepsilon\theta_t := \theta_{t+\varepsilon} - \theta_t$. In contrast to the standard, noncompact MSRJD field theory~\cite{MSRJD1973,HertzRoudiSollich2017}, the response field $k_t$ now takes values in $\mathbb{Z}$ rather than in $\mathbb{R}$. This integer structure comes directly from the Fourier sum in~\eqref{eq:dirac_comb} and is what keeps track of winding sectors.

For the It\^o discretization the Jacobian of the transformation from noise to phase increments is unity, so no extra determinant appears. As usual, this ensures causal response functions, a feature we use later when we analyze connected correlators in Sec.~\ref{sec:disorder_avg}.

\subsection{Villain resummation}
\label{sec:villain}

The sum over integers $k_t$ in~\eqref{eq:compact_msrjd} can be recast in a more convenient form using Poisson resummation. Define
\begin{equation}
y_t := \Delta_\varepsilon\theta_t - \varepsilon F(\theta_t,t),
\end{equation}
so that at fixed $y$ we have
\begin{equation}
\label{eq:k_sum}
\sum_{k\in\mathbb{Z}} \exp\big\{\ii k\,y - D\varepsilon\,k^2\big\}.
\end{equation}
Applying the Poisson summation formula
\begin{equation}
\sum_{k\in\mathbb{Z}} f(k) = \sum_{r\in\mathbb{Z}} \widehat{f}(2\pi r),
\end{equation}
with $f(k) = \exp(\ii ky - D\varepsilon k^2)$ and Fourier transform
\begin{equation}
\widehat{f}(\nu) = \sqrt{\frac{\pi}{D\varepsilon}}\exp\!\bigg[-\frac{(y-\nu)^2}{4D\varepsilon}\bigg],
\end{equation}
gives the Villain form~\cite{Villain1975,SiebererWachtelAltmanDiehl2016}:
\begin{equation}
\label{eq:villain_identity}
\sum_{k\in\mathbb{Z}} \ee^{\ii ky - D\varepsilon k^2}
= \sqrt{\frac{\pi}{D\varepsilon}} \sum_{r\in\mathbb{Z}} \exp\!\bigg[-\frac{(y - 2\pi r)^2}{4D\varepsilon}\bigg].
\end{equation}
The right-hand side is a sum of Gaussians centered at $y = 2\pi r$, i.e. a wrapped (periodic) Gaussian on the circle.

Using~\eqref{eq:villain_identity} slice by slice, the generating functional becomes
\begin{align}
\label{eq:villain_Z}
\mathcal{Z}[J] &\propto \int\!\mathcal{D}\theta \prod_t \sum_{r_t\in\mathbb{Z}} \exp\!\bigg[-\frac{\big(\Delta_\varepsilon\theta_t - \varepsilon F(\theta_t,t) - 2\pi r_t\big)^2}{4D\varepsilon}\bigg] \nonumber\\
&\qquad\qquad\qquad\qquad \times \ee^{\varepsilon\sum_t J_t O(\theta_t)}.
\end{align}
Each factor is a wrapped Gaussian in the increment, and $r_t\in\mathbb{Z}$ records the winding acquired during that time step. Taking the continuum limit $\varepsilon\to 0$ \emph{before} sending $D\to 0$ produces a compact field theory that still resolves phase slips. The ordering of these limits is important: it guarantees that rare winding events for small $D$ retain a finite weight in the path integral.

\subsection{Vector case: $N$ phases}
\label{sec:vector_case}

For $N$ coupled oscillators $\{\theta_{t,i}\}_{i=1}^N$, define
\begin{equation}
y_{t,i} := \Delta_\varepsilon\theta_{t,i} - \varepsilon F_i(\theta_t,t).
\end{equation}
The compact MSRJD action generalizes to
\begin{equation}
\label{eq:cmsrjd_N}
S_{\rm cMSRJD}[\theta,k] = \sum_{t,i}\Big[\ii k_{t,i}\,y_{t,i} - D\varepsilon\,k_{t,i}^2\Big], \qquad k_{t,i}\in\mathbb{Z},
\end{equation}
and the Villain representation reads
\begin{equation}
\label{eq:villain_N}
\prod_{t,i}\sum_{r_{t,i}\in\mathbb{Z}} \exp\!\bigg[-\frac{(y_{t,i} - 2\pi r_{t,i})^2}{4D\varepsilon}\bigg].
\end{equation}
This provides the starting point for including quenched coupling disorder, in close analogy with DMFT treatments of random neural networks and spin glasses~\cite{Sompolinsky1988,Crisanti2018,CrisantiSommers1993}.

\subsection{Composite fields and action decomposition}
\label{sec:composite_fields}

To connect this compact formulation to the coupling structure~\eqref{eq:coupling_split}, it is convenient to introduce composite fields built from the phase exponentials and the response variables:
\begin{equation}
\label{eq:Phi_Psi}
\Phi^{(m)}_{t,i} := \ee^{\ii m\theta_{t,i}}, \qquad \Psi^{(m)}_{t,i} := k_{t,i}\,\Phi^{(m)}_{t,i}.
\end{equation}
The harmonic order parameters~\eqref{eq:order_params} are simply $Z_m(t) = N^{-1}\sum_j \Phi^{(m)}_{t,j}$.

Substituting the coupling decomposition~\eqref{eq:coupling_split} into the drift~\eqref{eq:drift}, and using the harmonic expansion~\eqref{eq:harmonic}, the part of the MSRJD action that depends on the couplings splits into
\begin{equation}
S_W = S_{\rm mean} + S_{\rm rand},
\end{equation}
with the mean-channel contribution
\begin{equation}
\label{eq:S_mean}
S_{\rm mean} = -\ii\varepsilon \sum_{t,i}\sum_{m\ge 1}\Big(J_0 h_m\,\Psi^{(m)}_{t,i}\,\overline{Z_m(t)} + \text{c.c.}\Big),
\end{equation}
and the random-channel contribution
\begin{equation}
\label{eq:S_rand}
S_{\rm rand} = -\ii g\varepsilon \sum_{t,i,j}\sum_{m\ge 1}\Big(h_m\,\Psi^{(m)}_{t,i}\,\widetilde{W}_{ij}\,\overline{\Phi}^{(m)}_{t,j} + \text{c.c.}\Big).
\end{equation}
The mean channel couples the response composites $\Psi^{(m)}$ linearly to the collective variables $Z_m$, while the random channel involves the bilinear form $\Psi\,\widetilde{W}\,\overline{\Phi}$, which will generate time-nonlocal terms once we average over the quenched disorder. We turn to this average in the next section.

\section{Disorder Average and Emergence of Two-Time Fields}
\label{sec:disorder_avg}

We now average over the quenched random couplings $\widetilde{W}$ at fixed oscillator paths $\{\theta_{t,i}, k_{t,i}\}$. This produces time-nonlocal terms that link different time slices and naturally introduce two-time collective fields at the DMFT saddle point.

\subsection{Gaussian average over $\widetilde{W}$}
\label{sec:W_average}

The random couplings $\widetilde{W}_{ij}$ are independent and identically distributed (i.i.d.)\ with zero mean and variance $1/N$. Since the action~\eqref{eq:S_rand} is linear in $\widetilde{W}$, the disorder average of $\exp(S_{\rm rand})$ reduces to a Gaussian integral. Using
\begin{equation}
\big\langle \ee^{\sum_{ij} A_{ij}\widetilde{W}_{ij}}\big\rangle_{\widetilde{W}}
= \exp\!\bigg[\frac{1}{2N}\sum_{ij} A_{ij}^2\bigg]
\end{equation}
for arbitrary $A_{ij}$, we obtain
\begin{align}
\label{eq:W_avg_result}
\big\langle \ee^{S_{\rm rand}}\big\rangle_{\widetilde{W}}
&= \exp\!\bigg[\frac{g^2\varepsilon^2}{2N} \sum_{t,t'}\sum_{i,j}\sum_{m,n} \nonumber\\
&\quad \times \Big(h_m h_n^*\,\Psi^{(m)}_{t,i}\Psi^{(n)*}_{t',i}\,
\overline{\Phi}^{(m)}_{t,j}\Phi^{(n)}_{t',j} + \text{c.c.}\Big)\bigg].
\end{align}
The average thus generates a quartic interaction that is nonlocal in time: different times $t$ and $t'$ are coupled via products of $\Psi\Psi^*$ and $\overline{\Phi}\Phi$. In DMFT language, a single frozen realization of quenched disorder turns into self-generated temporal correlations in the $N\to\infty$ limit.

\subsection{Two-time collective fields}
\label{sec:two_time_fields}

To decouple the quartic term, we introduce two-time collective fields
\begin{align}
\label{eq:Q_def}
Q^{(m,n)}_{t,t'} &:= \frac{1}{N}\sum_{j=1}^N \overline{\Phi}^{(m)}_{t,j}\,\Phi^{(n)}_{t',j}, \\
\label{eq:S_def}
S^{(m,n)}_{t,t'} &:= \frac{1}{N}\sum_{i=1}^N \Psi^{(m)}_{t,i}\,\Psi^{(n)*}_{t',i}.
\end{align}
In terms of these fields, the exponent in~\eqref{eq:W_avg_result} can be written as
\begin{equation}
\frac{g^2\varepsilon^2 N}{2}\sum_{t,t',m,n}
\Big(h_m h_n^*\,S^{(m,n)}_{t,t'}\,Q^{(m,n)}_{t,t'} + \text{c.c.}\Big).
\end{equation}

The quantity $Q^{(m,n)}_{t,t'}$ is a circular two-time correlator of the phase exponentials at harmonics $m$ and $n$. In particular, $Q^{(m,m)}_{t,t} = 1$ because $|\ee^{\ii m\theta}|^2 = 1$. The field $S^{(m,n)}_{t,t'}$ is built from the response variables and will be seen to play a more limited role once causality is taken into account.

\subsection{Hubbard--Stratonovich decoupling}
\label{sec:HS}

To enforce the definitions~\eqref{eq:Q_def}--\eqref{eq:S_def}, we introduce conjugate fields $(\widehat{Q}, \widehat{S})$ using delta-function constraints represented as Fourier integrals. The partition function then takes the form
\begin{align}
\mathcal{Z} &\propto \int\!\mathcal{D}Q\,\mathcal{D}\widehat{Q}\,\mathcal{D}S\,\mathcal{D}\widehat{S}\;
\ee^{N\mathcal{S}_{\rm coll}[Q,\widehat{Q},S,\widehat{S}]} \nonumber\\
&\qquad \times \prod_{i=1}^N \mathcal{Z}^{(i)}_{\rm site}[Q,\widehat{Q},S,\widehat{S}],
\end{align}
with
\begin{align}
\label{eq:S_coll}
\mathcal{S}_{\rm coll}
&= \sum_{t,t',m,n}\bigg[
\widehat{Q}^{(m,n)}_{t,t'}\,Q^{(m,n)}_{t,t'}
+ \widehat{S}^{(m,n)}_{t,t'}\,S^{(m,n)}_{t,t'} \nonumber\\
&\quad + \frac{g^2\varepsilon^2}{2}
\Big(h_m h_n^*\,S^{(m,n)}_{t,t'}\,Q^{(m,n)}_{t,t'} + \text{c.c.}\Big)
\bigg],
\end{align}
and where $\mathcal{Z}^{(i)}_{\rm site}$ is the partition function of a single oscillator in the presence of the collective fields.

Stationarity with respect to $\widehat{Q}$ and $\widehat{S}$ yields
\begin{align}
\label{eq:Qhat_saddle}
\widehat{Q}^{(m,n)}_{t,t'}
&= -\frac{g^2\varepsilon^2}{2}\,h_m h_n^*\,S^{(m,n)}_{t,t'}, \\
\label{eq:Shat_saddle}
\widehat{S}^{(m,n)}_{t,t'}
&= -\frac{g^2\varepsilon^2}{2}\,h_m^* h_n\,Q^{(m,n)}_{t,t'},
\end{align}
so the two sectors are coupled through their conjugate fields.

\subsection{Causality and the $S$-channel}
\label{sec:causality}

The It\^o discretization in~\eqref{eq:discrete_update} implies that the response field $k_t$ at time $t$ depends only on the phase $\theta_t$ at the same time and not on future increments. Response correlators are therefore strictly causal and vanish for $t\neq t'$.

As a consequence, the two-time response correlator $S^{(m,n)}_{t,t'}$ defined in~\eqref{eq:S_def} only has equal-time contributions, $S^{(m,n)}_{t,t'}\propto \delta_{t,t'}$. These contact terms merely affect the overall normalization of $\mathcal{Z}$ and do not transmit information between distinct times. At the physical saddle point there is therefore no time-extended ($t\neq t'$) contribution from the $S$-channel.

The only genuinely dynamical two-time object that remains is $Q^{(m,n)}_{t,t'}$. The DMFT closure can thus be expressed entirely in terms of this correlator, which simplifies the structure compared with theories where both correlation and response have independent temporal dynamics.

\subsection{Colored Gaussian auxiliary field}
\label{sec:zeta_field}

With the $S$-sector effectively local in time, the residual bilinear dependence on $Q$ can be decoupled by introducing a complex Gaussian auxiliary field $\zeta^{(m)}_t$ with zero mean and covariance
\begin{equation}
\label{eq:zeta_cov_disc}
\big\langle \zeta^{(m)}_t\,\overline{\zeta}^{(n)}_{t'}\big\rangle
= g^2\,h_m h_n^*\,Q^{(m,n)}_{t,t'}.
\end{equation}
Integrating in $\zeta$ and collecting terms, the partition function factorizes over sites:
\begin{equation}
\mathcal{Z} \propto \int\!\mathcal{D}Q\;\ee^{N\mathcal{S}_{\rm eff}[Q]}
\prod_{i=1}^N \mathcal{Z}_{\rm site}[Q],
\end{equation}
where each factor $\mathcal{Z}_{\rm site}[Q]$ is the partition function of a single oscillator driven by the colored field $\zeta$ with covariance~\eqref{eq:zeta_cov_disc}, together with the deterministic mean-field contribution from $S_{\rm mean}$.

In the thermodynamic limit $N\to\infty$, the functional integral over $Q$ is dominated by its saddle point. The self-consistency condition at that point is
\begin{equation}
\label{eq:Q_selfconsist}
Q^{(m,n)}_{t,t'}
= \big\langle \overline{\Phi}^{(m)}_t\,\Phi^{(n)}_{t'}\big\rangle_{\rm 1\text{-}site}
= \big\langle \ee^{-\ii m\theta(t)}\,\ee^{\ii n\theta(t')}\big\rangle_{\rm 1\text{-}site},
\end{equation}
where the average is taken with respect to the single-site dynamics driven by $\zeta$ and the mean field. In other words, $Q$ fixes the statistics of $\zeta$, which drives the single-oscillator problem, whose solution must reproduce the same $Q$.

\subsection{Single-site weight}
\label{sec:site_weight}

Combining the mean-channel term~\eqref{eq:S_mean} with the colored-noise decoupling, the single-site partition function can be written as
\begin{align}
\label{eq:site_weight}
\log\mathcal{Z}_{\rm site}
&= \log\sum_{\{k_t\in\mathbb{Z}\}}\int\!\mathcal{D}\theta\;
\Big\langle\exp\Big\{\sum_t\big[\ii k_t(\Delta_\varepsilon\theta_t
- \varepsilon\omega - \varepsilon I_t) \nonumber\\
&\quad - D\varepsilon k_t^2\big]
- \ii\varepsilon\sum_{t,m}\big[k_t\Phi^{(m)}_t\overline{\zeta}^{(m)}_t + \text{c.c.}\big] \nonumber\\
&\quad - \ii\varepsilon\sum_{t,m}\big[J_0 h_m k_t\Phi^{(m)}_t\overline{Z_m(t)} + \text{c.c.}\big]\Big\}\Big\rangle_{\!\zeta}.
\end{align}
The first line describes the uncoupled phase dynamics with drift $\omega + I_t$ and diffusion $D$. The second line introduces the coupling to the colored Gaussian drive $\zeta$. The third line accounts for the deterministic mean field through the order parameters $Z_m$.

\subsection{From discrete weight to continuum SDE}
\label{sec:disc_to_cont}

To extract a continuum stochastic equation from~\eqref{eq:site_weight}, consider one time slice and combine the random and mean-field contributions into
\begin{equation}
J^{(m)}_t := \overline{\zeta}^{(m)}_t + J_0 h_m\,\overline{Z_m(t)}.
\end{equation}
After completing the square in the integer $k_t$ and using the Villain step as in Sec.~\ref{sec:villain}, the slice contributes a factor
\begin{equation}
\sum_{r\in\mathbb{Z}} \exp\!\bigg[-\frac{\big(y_t
- 2\varepsilon\,\mathrm{Re}\sum_m \Phi^{(m)}_t J^{(m)}_t - 2\pi r\big)^2}{4D\varepsilon}\bigg],
\end{equation}
where $y_t = \Delta_\varepsilon\theta_t - \varepsilon\omega - \varepsilon I_t$.

Taking the limit $\varepsilon\to 0$ turns the product of such wrapped Gaussians into the path measure of a continuous-time SDE on $S^1$. The effective single-site dynamics reads
\begin{align}
\label{eq:single_site_SDE}
\dot{\theta}(t)
&= \omega + I(t) + \sqrt{2D}\,\xi(t) \nonumber\\
&\quad + 2\sum_{m\ge 1}\mathrm{Re}\Big(J_0 h_m\,Z_m(t)\,\ee^{-\ii m\theta(t)}\Big) \nonumber\\
&\quad + 2\sum_{m\ge 1}\mathrm{Re}\Big(\zeta^{(m)}(t)\,\ee^{-\ii m\theta(t)}\Big),
\end{align}
where $\xi(t)$ is standard white noise and $\zeta^{(m)}(t)$ are complex Gaussian processes with
\begin{equation}
\label{eq:zeta_cov_cont}
\big\langle \zeta^{(m)}(t)\,\overline{\zeta}^{(n)}(t')\big\rangle
= g^2\,h_m h_n^*\,Q^{(m,n)}(t,t').
\end{equation}

The single-site equation~\eqref{eq:single_site_SDE} is the effective DMFT description: one oscillator on $S^1$ subject to intrinsic dynamics, a deterministic mean field encoded in $Z_m(t)$, and a colored Gaussian drive $\zeta^{(m)}$ with self-consistent covariance. The closure relations are
\begin{equation}
\label{eq:Z_closure}
\begin{split}
Z_m(t) &= \big\langle \ee^{\ii m\theta(t)} \big\rangle_{\rm 1\text{-}site},\\
Q^{(m,n)}(t,t')
&= \big\langle \ee^{-\ii m\theta(t)} \ee^{\ii n\theta(t')} \big\rangle_{\rm 1\text{-}site},
\end{split}
\end{equation}
where the averages are taken over realizations of the single-site SDE~\eqref{eq:single_site_SDE}.

For stationary, U(1)-invariant states with $Z_m = 0$ and time-translation invariance, the correlator depends only on the time difference: $Q^{(m,n)}(t,t') = \delta_{mn}Q_m(\tau)$ with $\tau = t - t'$ and $Q_m(0) = 1$. The next sections focus on this stationary regime.

\section{Mean-Channel Limit: Ott--Antonsen Reduction and Neural-Mass Models}
\label{sec:oa_limit}

The compact DMFT construction contains the standard mean-field picture as a special case. If we switch off the random channel by taking $g\to 0$, the colored drive disappears and only the deterministic coupling through the order parameters $Z_m$ remains. In this section we show how, under mild assumptions on the interaction, this limit reproduces the Ott--Antonsen (OA) manifold and recovers theta-neuron and quadratic integrate-and-fire (QIF) neural-mass equations.

\subsection{Mean-field dynamics on $S^1$}
\label{sec:mean_field_dynamics}

Setting $g=0$ in the single-site SDE~\eqref{eq:single_site_SDE} and taking $D=0$ gives the deterministic mean-field dynamics
\begin{equation}
\label{eq:mean_field_ode}
\dot{\theta} = F(\theta,t) + 2\sum_{m\ge 1}\mathrm{Re}\Big(J_0 h_m\,Z_m(t)\,\ee^{-\ii m\theta}\Big),
\end{equation}
where $F(\theta,t)$ collects the intrinsic frequency and any external input. The order parameters are defined self-consistently as
\begin{equation}
\label{eq:Z_mean_field}
Z_m(t) = \int_{-\pi}^{\pi}\!\ee^{\ii m\theta}\,f(\theta,\omega,t)\,d\theta\;p(\omega)\,d\omega,
\end{equation}
with $f(\theta,\omega,t)$ the phase density for oscillators of frequency $\omega$ and $p(\omega)$ the frequency distribution.

The density evolves according to the continuity equation on $S^1$,
\begin{equation}
\label{eq:continuity}
\partial_t f + \partial_\theta\big(v(\theta,\omega,t)\,f\big) = 0,
\end{equation}
with velocity field
\begin{align}
\label{eq:velocity}
v(\theta,\omega,t)
&= F(\theta,t) + J_0 h\,Z(t)\,\ee^{-\ii\theta}
   + J_0 h^*\,Z^*(t)\,\ee^{+\ii\theta} \nonumber\\
&= F(\theta,t) + 2J_0\,\mathrm{Re}\big(h\,Z(t)\,\ee^{-\ii\theta}\big),
\end{align}
where we have restricted to the first harmonic $m=1$.

\subsection{The Ott--Antonsen manifold}
\label{sec:OA_manifold}

Ott and Antonsen proposed that many mean-field oscillator systems are attracted to a low-dimensional manifold where the density takes the Poisson-kernel form~\cite{Ott2008,OttAntonsen2009}
\begin{equation}
\label{eq:OA_ansatz}
f(\theta,\omega,t)
= \frac{1}{2\pi}\,\frac{1 - |a(\omega,t)|^2}{
\big|1 - a(\omega,t)\,\ee^{-\ii\theta}\big|^2},
\end{equation}
with a complex amplitude $a(\omega,t)$ satisfying $|a|<1$. On this manifold, all Fourier modes of $f$ are powers of $a$:
\begin{equation}
\int_{-\pi}^{\pi}\!\ee^{\ii n\theta}\,f(\theta,\omega,t)\,d\theta
= \big(a(\omega,t)^*\big)^n, \qquad n\ge 1.
\end{equation}

For the OA manifold to be invariant under~\eqref{eq:continuity}, the velocity field must be at most first order in $\ee^{\pm\ii\theta}$. This is the case when the coupling is purely first harmonic ($h_m = 0$ for $m\ge 2$) and the intrinsic drift $F(\theta,t)$ is also linear in $\ee^{\pm\ii\theta}$.

Under these conditions, substituting~\eqref{eq:OA_ansatz} into~\eqref{eq:continuity} and using the algebra worked out in~\cite{Ott2008,OttAntonsen2009} leads to a closed equation for $a(\omega,t)$:
\begin{equation}
\label{eq:a_ode}
\dot{a}(\omega,t)
= -\ii\omega\,a + \frac{1}{2}\Big(H^*(t) - H(t)\,a^2\Big),
\end{equation}
with effective drive
\begin{equation}
\label{eq:H_eff}
H(t) = 2J_0 h\,Z(t) + 2I(t),
\end{equation}
and self-consistency condition
\begin{equation}
\label{eq:Z_from_a}
Z(t) = \int a^*(\omega,t)\,p(\omega)\,d\omega.
\end{equation}
Equations~\eqref{eq:a_ode}–\eqref{eq:Z_from_a} are the OA reduction: the PDE~\eqref{eq:continuity} is replaced by ODEs for $a(\omega,t)$ coupled through a single complex order parameter $Z(t)$.

\subsection{Lorentzian frequencies and residue closure}
\label{sec:lorentzian_closure}

For a Lorentzian frequency distribution~\eqref{eq:lorentzian} centered at $\omega_0$ with half-width $\Delta$, the integral in~\eqref{eq:Z_from_a} can be done analytically. The Lorentzian has a simple pole at $\omega = \omega_0 - \ii\Delta$ in the lower half-plane. If $a(\omega,t)$ is analytic there and decays for large $|\omega|$, we can close the contour and obtain
\begin{equation}
\label{eq:Z_residue}
Z(t) = a^*(\omega_0 - \ii\Delta, t).
\end{equation}

Evaluating~\eqref{eq:a_ode} at $\omega = \omega_0 - \ii\Delta$ and using~\eqref{eq:Z_residue} gives a single closed ODE for $Z$:
\begin{align}
\label{eq:Z_ode}
\dot{Z}
&= (\ii\omega_0 - \Delta)\,Z + \frac{1}{2}\big(H - H^*Z^2\big) \nonumber\\
&= (\ii\omega_0 - \Delta)\,Z + (J_0 h\,Z + I)
   - (J_0 h^*Z^* + I)\,Z^2.
\end{align}
This is the standard OA equation for a mean-field ensemble with Lorentzian frequency spread. The linear term $-\Delta Z$ represents dephasing from heterogeneity, and the nonlinear terms encode the feedback through $Z$.

\subsection{Kuramoto model}
\label{sec:kuramoto}

The classical Kuramoto model corresponds to $H(\Delta) = \sin\Delta$, i.e.\ $h = -\ii/2$ and $h^* = +\ii/2$. With $I=0$, Eq.~\eqref{eq:Z_ode} reduces to
\begin{equation}
\label{eq:kuramoto_ode}
\dot{Z}
= (\ii\omega_0 - \Delta)\,Z + \frac{J_0}{2}\big(Z - Z^*Z^2\big).
\end{equation}
Writing $Z = R\,\ee^{\ii\psi}$ with $R = |Z|$, the radial dynamics is
\begin{equation}
\label{eq:R_ode}
\dot{R} = -\Delta R + \frac{J_0}{2}\,R(1 - R^2).
\end{equation}
The incoherent state $R=0$ is stable for $J_0 < J_{0,c} = 2\Delta$ and unstable otherwise, and for $J_0 > 2\Delta$ there is a synchronized branch
\begin{equation}
R = \sqrt{1 - \frac{2\Delta}{J_0}}.
\end{equation}
This reproduces the classical Kuramoto transition~\cite{Kuramoto1975,Strogatz2000,Acebron2005}.

\subsection{Theta-neuron and QIF neural-mass models}
\label{sec:theta_QIF}

The theta-neuron model describes a type-I excitable neuron via a phase variable $\theta\in S^1$, with membrane voltage given by a stereographic transform of $\theta$. The intrinsic dynamics is
\begin{equation}
\label{eq:theta_neuron}
\dot{\theta}
= 1 - \cos\theta + (1 + \cos\theta)\,\eta,
\end{equation}
where $\eta$ controls excitability, and spikes correspond to $\theta$ crossing $\pi$.

For a population with excitabilities $\eta \sim p(\eta)$ and mean-field synaptic coupling, one typically obtains
\begin{equation}
F(\theta,t)
= 1 - \cos\theta + (1 + \cos\theta)\big(\eta + I(t)\big),
\end{equation}
where $I(t)$ collects external input and recurrent drive proportional to the population firing rate.

This drift is linear in $\ee^{\pm\ii\theta}$:
\begin{equation}
F(\theta,t)
= (1 + \eta + I)
  - \frac{1}{2}(\eta + I - 1)\ee^{\ii\theta}
  - \frac{1}{2}(\eta + I - 1)\ee^{-\ii\theta},
\end{equation}
so the OA ansatz applies. Combined with first-harmonic coupling, the flow remains M\"obius-invariant, and OA yields a closed two-dimensional system.

For Lorentzian-distributed excitabilities with center $\bar{\eta}$ and half-width $\Delta_\eta$, the residue step leads to the neural-mass equations of Montbri\'o, Paz\'o, and Roxin~\cite{Montbrio2015}. Introducing macroscopic variables $(r,v)$ via
\begin{equation}
Z = \frac{1 - W^*}{1 + W^*}, \qquad W = \pi r + \ii v,
\end{equation}
the OA equation~\eqref{eq:Z_ode} can be rewritten as
\begin{align}
\label{eq:QIF_mass}
\dot{r} &= \frac{\Delta_\eta}{\pi} + 2rv, \\
\dot{v} &= v^2 + \bar{\eta} + I(t) + J_{\rm syn}\,r - \pi^2 r^2,
\end{align}
where $r(t)$ is the population firing rate and $v(t)$ the mean membrane potential. These equations are exact in the limit of infinitely many all-to-all coupled QIF neurons~\cite{Montbrio2015,PazoMontbrio2014,Laing2014,BickReview2020}.

\subsection{General coupling and the role of $h$}
\label{sec:general_h}

The OA reduction also covers general first-harmonic interactions
\begin{equation}
H(\Delta)
= h\,\ee^{\ii\Delta} + h^*\ee^{-\ii\Delta},
\qquad h = |h|\ee^{\ii\alpha}.
\end{equation}
The synchronization threshold then depends on both $|h|$ and the phase $\alpha$. For Lorentzian frequencies, linearizing~\eqref{eq:Z_ode} around $Z=0$ gives the critical coupling
\begin{equation}
\label{eq:J0c_general}
J_{0,c} = \frac{\Delta}{|h|\,s},
\qquad s := -\sin\alpha = -\sin(\arg h),
\end{equation}
as long as $s>0$ (i.e.\ $\alpha \in (-\pi,0)$). The factor $s$ picks out the synchronizing component of the interaction: pure cosine coupling ($\alpha = 0$ or $\pi$) has $s=0$ and cannot create a synchronized state, while pure sine coupling ($\alpha = -\pi/2$) gives $s=1$ and the strongest tendency to synchronize.

In Sec.~\ref{sec:iprc} we show that, for phase-reduced neuron models, the coefficient $h$ obtained from the iPRC typically has a nonzero phase $\alpha$. As a result, the synchronization threshold can differ substantially from the textbook Kuramoto value $J_{0,c} = 2\Delta$.

\section{From Biophysical Neurons to Network Dynamics}
\label{sec:iprc}

The compact DMFT formulation treats the interaction function $H(\Delta)$ only through its Fourier coefficients~\eqref{eq:harmonic}, so any $2\pi$-periodic coupling can in principle be used. For neuron models with a stable limit cycle, phase reduction provides a concrete way to obtain these coefficients from the infinitesimal phase response curve (iPRC). This gives a direct route from single-cell biophysics to network-level DMFT predictions.

\subsection{Phase reduction and the infinitesimal PRC}
\label{sec:phase_reduction}

Consider a conductance-based or integrate-and-fire neuron model whose dynamics admits a stable limit cycle $\gamma$ with period $T$. Phase reduction~\cite{BrownMoehlisHolmes2004,ErmentroutTerman2010,Winfree1967} associates to each point on $\gamma$ a phase $\phi \in [0,2\pi)$ that advances uniformly in time:
\begin{equation}
\dot{\phi} = \omega_0 := \frac{2\pi}{T}
\end{equation}
in the absence of perturbations.

Under a weak perturbation $\epsilon\,p(t)$, the phase obeys
\begin{equation}
\label{eq:phase_reduction}
\dot{\phi} = \omega_0 + \epsilon\,Z(\phi)\,p(t) + O(\epsilon^2),
\end{equation}
where $Z(\phi)$ is the iPRC. The function $Z$ quantifies how a brief input delivered at phase $\phi$ shifts subsequent spike times: $Z(\phi)>0$ corresponds to an advance, $Z(\phi)<0$ to a delay.

The iPRC is a property of the single neuron. It can be obtained numerically from the limit cycle via the adjoint method, or measured experimentally by injecting brief current pulses at different phases and recording the induced phase shifts~\cite{NeteaCMNS2012,SchultheissPRE2010}.

\subsection{From iPRC to coupling coefficients}
\label{sec:iprc_to_h}

Now consider a network of $N$ identical neurons with weak synaptic coupling. After phase reduction, the network dynamics can be written in the form~\eqref{eq:model}–\eqref{eq:drift}, with an interaction function $H(\Delta\phi)$ that depends on both the iPRC and the synaptic waveform.

For instantaneous, $\delta$-like synapses with amplitude proportional to a postsynaptic conductance change, one effectively samples the iPRC at the postsynaptic phase:
\begin{equation}
H(\Delta\phi) \propto Z(\phi_{\rm post}),
\qquad
\Delta\phi = \phi_{\rm post} - \phi_{\rm pre}.
\end{equation}
For synapses with a finite time course $s(t)$, the interaction is given by a convolution of $Z$ with the synaptic kernel, but the resulting $H(\Delta\phi)$ is still $2\pi$-periodic and can be expanded in Fourier modes~\cite{BrownMoehlisHolmes2004,ErmentroutKopell1986}.

Given a numerically computed or experimentally measured iPRC $Z(\phi)$, we obtain the Fourier coefficients via
\begin{equation}
\label{eq:h_from_Z}
h_m = \frac{1}{2\pi}\int_0^{2\pi} Z(\phi)\,\ee^{-\ii m\phi}\,d\phi,
\qquad m \ge 1.
\end{equation}
For a real-valued iPRC, $h_{-m} = h_m^*$, and the interaction function can be written as
\begin{align}
H(\Delta\phi)
&= \sum_{m\ge 1}\bigl(h_m\ee^{\ii m\Delta\phi}
    + h_m^*\ee^{-\ii m\Delta\phi}\bigr) \nonumber\\
&= 2\sum_{m\ge 1}\mathrm{Re}\!\bigl(h_m\ee^{\ii m\Delta\phi}\bigr).
\end{align}
Once the set $\{h_m\}$ is known, it can be plugged into the DMFT equations. The mean-channel dynamics of Sec.~\ref{sec:oa_limit} then follows without further reference to the underlying biophysical model.

\subsection{Example: Adaptive exponential integrate-and-fire neuron}
\label{sec:adex}

As a concrete example we consider the adaptive exponential integrate-and-fire (AdEx) model~\cite{Brette2005}, a two-dimensional point-neuron model that captures spike initiation, adaptation, and subthreshold dynamics. Its equations are
\begin{align}
C \dot{V} &= -g_L(V - E_L)
          + g_L\Delta_T \exp\!\Big(\frac{V - V_T}{\Delta_T}\Big)
          - w + I, \\
\tau_w \dot{w} &= a(V - E_L) - w,
\end{align}
with reset $V\to V_r$ and $w\to w + b$ when $V$ reaches a cutoff. For suitable parameters, the dynamics converges to a stable limit cycle corresponding to tonic firing.

We compute the iPRC using the adjoint method~\cite{ErmentroutTerman2010,BrownMoehlisHolmes2004}. Linearizing the dynamics around the limit cycle and solving the adjoint variational equation yields a $2\pi$-periodic function $Z(\phi)$, normalized so that
\begin{equation}
\int_0^{2\pi} Z(\phi)\,\nabla_\phi \gamma\,d\phi = 1,
\end{equation}
where $\gamma$ parametrizes the limit cycle. A representative iPRC is shown in Fig.~\ref{fig:iprc}(a). The curve is strongly non-sinusoidal: it displays a pronounced peak near the upstroke of the spike and small amplitude during the refractory part of the cycle.

From this iPRC we extract the first Fourier coefficient via~\eqref{eq:h_from_Z},
\begin{equation}
h_1 = |h_1|\,\ee^{\ii\alpha},
\end{equation}
with magnitude $|h_1|$ and phase $\alpha$. For the parameter set used in Fig.~\ref{fig:iprc}, we obtain $\alpha \approx -0.98$ rad, corresponding to a synchronizing component $s = -\sin\alpha \approx 0.83$. This value lies between pure sine coupling ($\alpha = -\pi/2$, $s=1$) and pure cosine coupling ($\alpha = 0$, $s=0$), and reflects the detailed shape of the AdEx iPRC.

\subsection{Mean-channel synchronization: theory versus simulation}
\label{sec:adex_sync}

With the iPRC-derived coefficient $h_1$ in hand, the OA reduction of Sec.~\ref{sec:oa_limit} predicts a synchronization threshold given by~\eqref{eq:J0c_general}:
\begin{equation}
J_{0,c}^{\rm th} = \frac{\Delta}{|h_1|\,s},
\end{equation}
where $\Delta$ is the half-width of the Lorentzian frequency distribution. For $J_0 > J_{0,c}^{\rm th}$ the OA dynamics yields
\begin{equation}
\label{eq:R_above_threshold}
R = |Z| = \sqrt{1 - \frac{J_{0,c}^{\rm th}}{J_0}},
\end{equation}
while $R=0$ for $J_0 \le J_{0,c}^{\rm th}$.

To check this prediction, we simulate a network of $N=2000$ phase oscillators with interaction
\begin{equation}
H(\Delta\phi) = 2\,\mathrm{Re}\bigl(h_1\ee^{\ii\Delta\phi}\bigr),
\end{equation}
using the AdEx-derived $h_1$. Intrinsic frequencies are drawn from a Lorentzian with half-width $\Delta = 0.1$. For each coupling $J_0$ we evolve the system to steady state and measure the time-averaged order parameter
\begin{equation}
R = \big\langle |Z_1(t)|\big\rangle_t.
\end{equation}

\begin{figure}[t]
\centering
\includegraphics[width=\columnwidth]{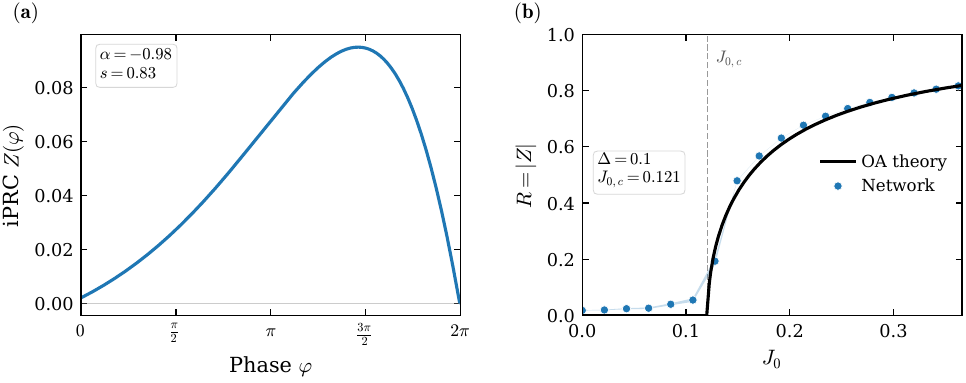}
\caption{iPRC-based parameterization of compact DMFT.
(a) Infinitesimal phase response curve $Z(\phi)$ for an AdEx neuron, computed via the adjoint method. The first Fourier harmonic has phase $\alpha \approx -0.98$ rad and synchronizing component $s = -\sin\alpha \approx 0.83$.
(b) Mean-channel synchronization: order parameter $R = |Z|$ versus coupling strength $J_0$ for a network of $N = 2000$ phase oscillators with AdEx-derived coupling. Blue circles: network simulations (mean $\pm$ SEM across runs). Black curve: OA prediction~\eqref{eq:R_above_threshold}. Dashed line: theoretical threshold $J_{0,c}^{\rm th} = \Delta/(|h_1| s) \approx 0.121$. Parameters: $\Delta = 0.1$, $D = 0$.}
\label{fig:iprc}
\end{figure}

The comparison is shown in Fig.~\ref{fig:iprc}(b). The simulation points follow the OA prediction closely, and the transition occurs near the theoretical threshold $J_{0,c}^{\rm th} \approx 0.121$. This agreement indicates that the iPRC–DMFT pipeline reproduces the collective synchronization behavior of the phase-reduced network quantitatively.

\subsection{Scope and relation to prior work}
\label{sec:iprc_scope}

The iPRC–DMFT construction presented here is complementary to voltage-based mean-field approaches.

A large body of work derives mean-field equations for AdEx and related models directly in voltage space~\cite{Augustin2017,Zerlaut2017,diVolo2019,Carlu2020}. There one starts from the Fokker–Planck equation for the membrane potential distribution, uses diffusion or quasi-renewal approximations to obtain a transfer function linking firing rate to input statistics, and finally closes on low-dimensional equations for rate and adaptation. This strategy preserves subthreshold voltage statistics and the full spike-generation mechanism, but it requires solving high-dimensional Fokker–Planck or integral equations for each parameter regime.

Our route is different. We first reduce the neuron to a phase oscillator, compute the iPRC once to obtain the coefficients $\{h_m\}$, and then work entirely within the DMFT framework. The price is that phase reduction assumes operation near a stable limit cycle and discards subthreshold statistics. The benefit is a compact analytic description: synchronization thresholds and other mean-field properties follow from a small set of Fourier coefficients.

The QIF neuron and its theta-neuron representation~\cite{Montbrio2015,PazoMontbrio2014,Laing2014} can be viewed as a special case in which phase reduction is exact, the iPRC is known in closed form, and OA gives exact mean-field equations. The present framework extends this picture to more detailed neuron models, where the effective coupling is obtained numerically from the iPRC rather than imposed \emph{a priori}.

\section{Numerical Validation}
\label{sec:numerics}

We now compare the compact DMFT predictions with numerical simulations of both the full $N$-oscillator system and the effective single-site description. Throughout this section we restrict to the first harmonic ($m=1$) with Kuramoto-type coupling $H(\Delta) = \sin\Delta$, so that $h = -\ii/2$ and $|h| = 1/2$.

\subsection{Simulation methodology}
\label{sec:methodology}

For network simulations, we integrate the $N$-body dynamics~\eqref{eq:model} with coupling matrix~\eqref{eq:coupling_split} using an Euler--Maruyama scheme with time step $dt = 0.01$. The random matrix $\widetilde{W}$ is drawn once per realization with i.i.d.\ entries $\widetilde{W}_{ij}\sim\mathcal{N}(0,1/N)$. Intrinsic frequencies are sampled from a Lorentzian with half-width $\Delta$. After discarding an initial transient (typically $T_{\rm trans} = 200$ time units), we estimate the circular two-time correlator
\begin{equation}
Q(\tau)
  = \frac{1}{N}\sum_{j=1}^N
      \big\langle \ee^{-\ii\theta_j(t)}\,\ee^{\ii\theta_j(t+\tau)}\big\rangle_t
\end{equation}
by averaging over time.

To test the DMFT closure, we also simulate the single-site SDE~\eqref{eq:single_site_SDE} driven by a synthetic colored noise $\zeta(t)$ with prescribed covariance determined by a candidate $Q(\tau)$. Given $Q(\tau)$, we compute its power spectrum $\widehat{Q}(\Omega)$, generate complex Gaussian white noise in frequency space, multiply by $\sqrt{\widehat{Q}(\Omega)}$, and transform back to obtain $\zeta(t)$. This construction yields a process with the desired two-point statistics.

Starting from an initial guess $Q^{(0)}(\tau)$, usually the incoherent form $Q^{(0)}(\tau) = \ee^{-\gamma_0|\tau|}$ with $\gamma_0$ measured at $g=0$, we iterate as follows. Given $Q^{(n)}$, we synthesize $\zeta$, simulate the single-site SDE for multiple independent realizations, measure the resulting correlator $Q^{(n+1)}_{\rm new}(\tau)$, and update
\begin{equation}
Q^{(n+1)} \leftarrow \alpha\,Q^{(n+1)}_{\rm new} + (1-\alpha)\,Q^{(n)},
\qquad
\alpha \in [0.3,0.5].
\end{equation}
We declare convergence when $\|Q^{(n+1)} - Q^{(n)}\|_\infty < 10^{-3}$.

\subsection{DMFT correlator closure}
\label{sec:surrogate}

\begin{figure}[t]
\centering
\includegraphics[width=\columnwidth]{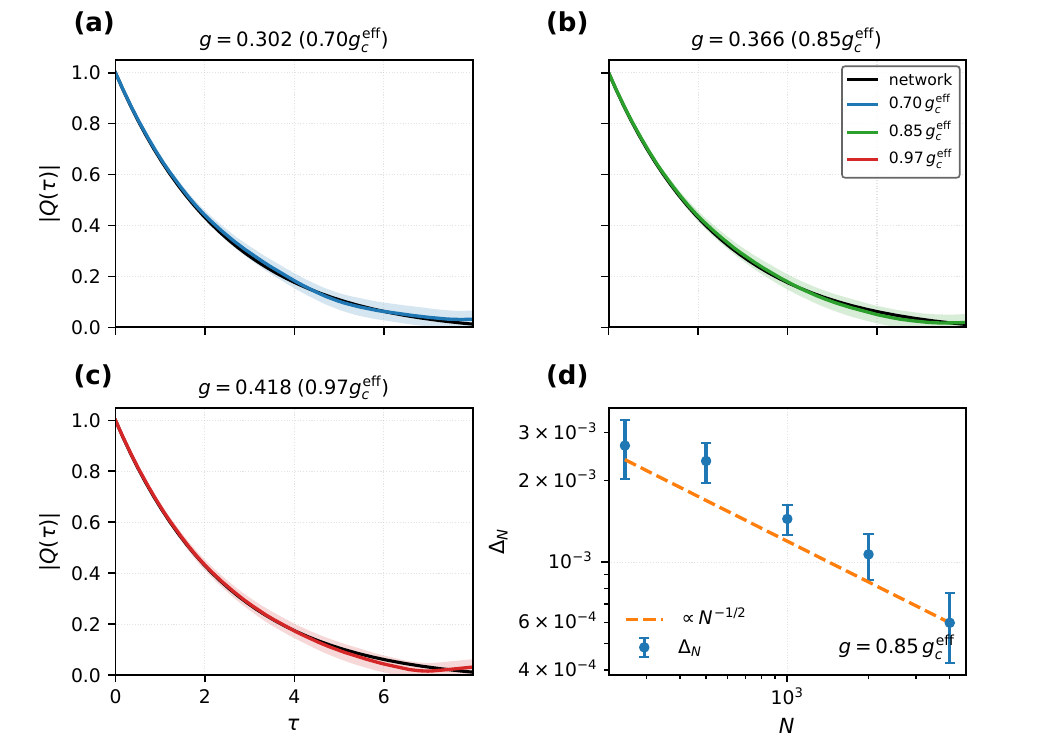}
\caption{DMFT correlator closure and finite-size scaling.
(a--c) Magnitude $|Q(\tau)|$ from direct network simulations (black curves, mean $\pm$ SEM across probes and disorder realizations) compared with single-rotor DMFT dynamics driven by self-consistent colored noise (shaded bands, mean $\pm$ SEM across independent noise draws) at three subcritical couplings $g/g_c^{\rm eff} \in \{0.70, 0.85, 0.97\}$. Here $g_c^{\rm eff} = \gamma_0/|h|$ is set by the measured incoherent dephasing rate.
(d) Finite-size deviation $\Delta_N$ at $g = 0.85\,g_c^{\rm eff}$ scales as $N^{-1/2}$, consistent with central-limit fluctuations around the DMFT limit. Parameters: $D = 0.05$, $\Delta = 0.3$, Lorentzian $p(\omega)$.}
\label{fig:surrogate}
\end{figure}

The self-consistent construction~\eqref{eq:single_site_SDE}–\eqref{eq:Z_closure} replaces the dense $N$-body system by a single rotor driven by colored noise whose covariance is fixed by $Q(\tau)$. We assess this reduction by directly comparing $Q(\tau)$ measured in full networks with $Q(\tau)$ obtained from the single-site dynamics driven by synthesized $\zeta$.

To set an overall scale, we first estimate the incoherent dephasing rate $\gamma_0$ from simulations at $g=0$, by fitting $|Q(\tau)| \sim \ee^{-\gamma_0 \tau}$ over an intermediate time window. We then define an effective critical scale $g_c^{\rm eff} = \gamma_0/|h|$. Figure~\ref{fig:surrogate}(a--c) shows $|Q(\tau)|$ for three subcritical couplings $g/g_c^{\rm eff}\in\{0.70, 0.85, 0.97\}$. In each panel, the DMFT surrogate (colored band) closely follows the network result (black curve).

The agreement improves as $g$ approaches $g_c^{\rm eff}$: at $g = 0.70\,g_c^{\rm eff}$ small discrepancies are visible at intermediate lags, whereas at $g = 0.97\,g_c^{\rm eff}$ the two curves are nearly indistinguishable on the plotted scale. This is consistent with the idea that, in the dense, strongly disordered regime, the effective colored drive is well described by the DMFT self-consistency.

We also examine how deviations from the DMFT prediction scale with system size. For each $N$ we define
\begin{equation}
\Delta_N
  = \bigg[
      \frac{1}{\tau_{\max}}
      \int_0^{\tau_{\max}}
        \big|Q_N(\tau) - Q_{\rm ref}(\tau)\big|^2\,d\tau
    \bigg]^{1/2},
\end{equation}
where $Q_{\rm ref}(\tau)$ is a reference correlator obtained from the largest simulated system. At $g = 0.85\,g_c^{\rm eff}$ we find $\Delta_N \propto N^{-1/2}$ over $N\in[250,4000]$ [Fig.~\ref{fig:surrogate}(d)], as expected if the effective field is a sum of $O(N)$ weakly correlated contributions.

We emphasize that this validation probes the dense-coupling regime in which the present DMFT closure is expected to be most accurate, where each oscillator receives a sum of $O(N)$ random inputs of size $O(N^{-1/2})$, so the disorder-induced local field approaches a Gaussian colored process in the thermodynamic limit. Moving away from this regime can introduce two distinct sources of discrepancy. First, for weak disorder ($g \ll g_c^{\rm eff}$) the disorder-driven fluctuations are small, so finite-size effects and statistical uncertainty can be comparatively more visible relative to the size of the random-channel contribution, since in this regime the random channel acts only as a small perturbation about the mean-field dynamics. Second, for sparse graphs with finite mean degree, the local field is no longer a sum over $O(N)$ weak terms and need not be well-approximated as Gaussian, and extending the compact DMFT to that setting would likely require a sparse dynamical mean-field formulation that tracks a distribution over effective field trajectories rather than a single correlator $Q(\tau)$~\cite{HatchettUezu2008,Metz2025SparseDMFT}.

Taken together, these tests indicate that, for dense random coupling, the compact DMFT captures the two-time statistics of the oscillator phases. The single-site equation~\eqref{eq:single_site_SDE} with colored noise covariance fixed by $Q(\tau)$ provides a quantitatively accurate reduced description of the $N$-body dynamics in the thermodynamic limit.

\section{Discussion}
\label{sec:conclusion}

We have formulated a compact dynamical mean-field theory for networks of phase oscillators on $S^1$ that keeps the circle topology explicit in the path-integral description. Enforcing $2\pi$-periodicity via a Dirac-comb constraint and using Villain resummation leads, after disorder averaging, to a self-consistent single-oscillator SDE driven by a deterministic mean field and colored Gaussian input with covariance fixed by a circular two-time correlator $Q(\tau)$. Splitting the coupling into coherent and disordered parts interpolates between classical mean-field synchronization ($g\to 0$, where the Ott--Antonsen and Kuramoto/theta-neuron reductions are recovered) and disorder-dominated regimes, where the DMFT closure captures the statistics of fully disordered networks and agrees with direct simulations.

A practical advantage of the framework is that it accepts arbitrary $2\pi$-periodic coupling functions through their Fourier coefficients. For neuron models with a stable limit cycle, these coefficients can be obtained directly from the infinitesimal phase response curve, giving a concrete route from single-cell dynamics to network-level mean-field predictions. We illustrated this for AdEx neurons and found quantitative agreement between iPRC-based mean-field theory and simulations of phase-reduced networks. Future work can include extensions to sparse or structured connectivity, where a distributional sparse-graph cavity/DMFT closure may be required~\cite{HatchettUezu2008,Metz2025SparseDMFT}, spatially extended systems such as one-dimensional rings and two-dimensional neural fields, and regimes with intrinsically non-Gaussian fluctuation statistics.

The simulation code supporting Figs.~1 and~2 is openly available in Ref.~\cite{Reddy_Zenodo_2026}.

\bibliography{apssamp}

@PREAMBLE{
 "\providecommand{\noopsort}[1]{}"

}

@incollection{Kuramoto1975,
  author    = {Yoshiki Kuramoto},
  title     = {Self-entrainment of a population of coupled nonlinear oscillators},
  booktitle = {International Symposium on Mathematical Problems in Theoretical Physics},
  editor    = {H. Araki},
  series    = {Lecture Notes in Physics},
  volume    = {39},
  pages     = {420--422},
  publisher = {Springer},
  address   = {Berlin},
  year      = {1975},
  doi       = {10.1007/BFb0013365}
}

@article{Strogatz2000,
  author  = {Steven H. Strogatz},
  title   = {From {Kuramoto} to {Crawford}: exploring the onset of synchronization in populations of coupled oscillators},
  journal = {Physica D},
  volume  = {143},
  number  = {1-4},
  pages   = {1--20},
  year    = {2000},
  doi     = {10.1016/S0167-2789(00)00094-4}
}

@article{Acebron2005,
  author = {Juan A. Acebr{\'o}n and Luis L. Bonilla and 
          Conrad J. P{\'e}rez Vicente and F{\'e}lix Ritort and Renato Spigler},
  title   = {The {Kuramoto} model: A simple paradigm for synchronization phenomena},
  journal = {Rev. Mod. Phys.},
  volume  = {77},
  pages   = {137--185},
  year    = {2005},
  doi     = {10.1103/RevModPhys.77.137}
}

@article{Rodrigues2016,
  author  = {Francisco A. Rodrigues and Thomas K. D. M. Peron and Peng Ji and J{\"u}rgen Kurths},
  title   = {The {Kuramoto} model in complex networks},
  journal = {Phys. Rep.},
  volume  = {610},
  pages   = {1--98},
  year    = {2016},
  doi     = {10.1016/j.physrep.2015.10.008}
}

@article{Ott2008,
  author  = {Edward Ott and Thomas M. Antonsen},
  title   = {Low dimensional behavior of large systems of globally coupled oscillators},
  journal = {Chaos},
  volume  = {18},
  number  = {3},
  pages   = {037113},
  year    = {2008},
  doi     = {10.1063/1.2930766}
}

@article{Watanabe1993,
  author  = {Shinya Watanabe and Steven H. Strogatz},
  title   = {Integrability of a globally coupled oscillator array},
  journal = {Phys. Rev. Lett.},
  volume  = {70},
  number  = {16},
  pages   = {2391--2394},
  year    = {1993},
  doi     = {10.1103/PhysRevLett.70.2391}
}

@article{OttAntonsen2009,
  author  = {Edward Ott and Thomas M. Antonsen},
  title   = {Long time evolution of phase oscillator systems},
  journal = {Chaos},
  volume  = {19},
  number  = {2},
  pages   = {023117},
  year    = {2009},
  doi     = {10.1063/1.3136851}
}

@article{BickReview2020,
  author  = {Christian Bick and Marc Goodfellow and 
           Carlo R. Laing and Erik A. Martens},
  title   = {Understanding the dynamics of biological and neural oscillator networks through exact mean-field reductions: A review},
  journal = {J. Math. Neurosci.},
  volume  = {10},
  number  = {1},
  pages   = {9},
  year    = {2020},
  doi     = {10.1186/s13408-020-00086-9}
}

@article{Montbrio2015,
  author  = {Ernest Montbri{\'o} and Diego Paz{\'o} and Alex Roxin},
  title   = {Macroscopic description for networks of spiking neurons},
  journal = {Phys. Rev. X},
  volume  = {5},
  number  = {2},
  pages   = {021028},
  year    = {2015},
  doi     = {10.1103/PhysRevX.5.021028}
}

@article{PazoMontbrio2014,
  author  = {Diego Paz{\'o} and Ernest Montbri{\'o}},
  title   = {Low-dimensional dynamics of populations of pulse-coupled oscillators},
  journal = {Phys. Rev. X},
  volume  = {4},
  number  = {1},
  pages   = {011009},
  year    = {2014},
  doi     = {10.1103/PhysRevX.4.011009}
}

@article{Laing2014,
  author  = {Carlo R. Laing},
  title   = {Derivation of a neural field model from a network of theta neurons},
  journal = {Phys. Rev. E},
  volume  = {90},
  number  = {1},
  pages   = {010901},
  year    = {2014},
  doi     = {10.1103/PhysRevE.90.010901}
}

@article{Sompolinsky1988,
  author  = {Haim Sompolinsky and Andrea Crisanti and Hans-J{\"u}rgen Sommers},
  title   = {Chaos in Random Neural Networks},
  journal = {Physical Review Letters},
  volume  = {61},
  number  = {3},
  pages   = {259--262},
  year    = {1988},
  doi     = {10.1103/PhysRevLett.61.259}
}

@article{Crisanti2018,
  author  = {Andrea Crisanti and Haim Sompolinsky},
  title   = {Path-integral approach to random neural networks},
  journal = {Phys. Rev. E},
  volume  = {98},
  number  = {6},
  pages   = {062120},
  year    = {2018},
  doi     = {10.1103/PhysRevE.98.062120}
}

@article{CrisantiSommers1993,
  author  = {Andrea Crisanti and Heinz Horner and Hans-J{\"u}rgen Sommers},
  title   = {The spherical {$p$}-spin interaction spin-glass model},
  journal = {Z. Phys. B},
  volume  = {92},
  pages   = {257--271},
  year    = {1993},
  doi     = {10.1007/BF01312184}
}

@article{HertzRoudiSollich2017,
  author  = {John A. Hertz and Yasser Roudi and Peter Sollich},
  title   = {Path integral methods for the dynamics of stochastic and disordered systems},
  journal = {J. Phys. A: Math. Theor.},
  volume  = {50},
  number  = {3},
  pages   = {033001},
  year    = {2017},
  doi     = {10.1088/1751-8121/50/3/033001}
}

@article{MSRJD1973,
  author  = {Paul C. Martin and E. D. Siggia and H. A. Rose},
  title   = {Statistical Dynamics of Classical Systems},
  journal = {Physical Review A},
  volume  = {8},
  number  = {1},
  pages   = {423--437},
  year    = {1973},
  doi     = {10.1103/PhysRevA.8.423}
}

@article{Villain1975,
  author  = {Jacques Villain},
  title   = {Theory of one- and two-dimensional magnets with an easy magnetization plane. II. The planar, classical, two-dimensional magnet},
  journal = {Journal de Physique},
  volume  = {36},
  number  = {6},
  pages   = {581--590},
  year    = {1975},
  doi     = {10.1051/jphys:01975003606058100}
}

@article{Prueser2024PRL,
  author  = {Axel Pr{\"u}ser and Sebastian Rosmej and Andreas Engel},
  title   = {Nature of the Volcano Transition in the Fully Disordered Kuramoto Model},
  journal = {Physical Review Letters},
  volume  = {132},
  number  = {18},
  pages   = {187201},
  year    = {2024},
  doi     = {10.1103/PhysRevLett.132.187201}
}

@article{PrueserEngel2024PRE,
  author  = {Axel Pr{\"u}ser and Andreas Engel},
  title   = {Role of Coupling Asymmetry in the Fully Disordered Kuramoto Model},
  journal = {Physical Review E},
  volume  = {110},
  number  = {6},
  pages   = {064214},
  year    = {2024},
  doi     = {10.1103/PhysRevE.110.064214}
}

@article{PazoGallego2023PRE,
  author  = {Diego Paz{\'o} and Rafael Gallego},
  title   = {Volcano transition in populations of phase oscillators with random nonreciprocal interactions},
  journal = {Physical Review E},
  volume  = {108},
  number  = {1},
  pages   = {014202},
  year    = {2023},
  doi     = {10.1103/PhysRevE.108.014202}
}

@article{KatiRanftLindner2024PRE,
  author  = {Yagmur Kati and Jonas Ranft and Benjamin Lindner},
  title   = {Self-consistent autocorrelation of a disordered Kuramoto model in the asynchronous state},
  journal = {Physical Review E},
  volume  = {110},
  number  = {5},
  pages   = {054301},
  year    = {2024},
  doi     = {10.1103/PhysRevE.110.054301}
}

@article{OttinoLoefflerStrogatz2018PRL,
  author = {Bertrand Ottino-L{\"o}ffler and Steven H. Strogatz},
  title   = {Volcano Transition in a Solvable Model of Frustrated Oscillators},
  journal = {Physical Review Letters},
  volume  = {120},
  number  = {26},
  pages   = {264102},
  year    = {2018},
  doi     = {10.1103/PhysRevLett.120.264102}
}

@article{LeonPazo2025Chaos,
  author  = {Iv{\'a}n Le{\'o}n and Diego Paz{\'o}},
  title   = {Dynamics and chaotic properties of the fully disordered {Kuramoto} model},
  journal = {Chaos},
  volume  = {35},
  number  = {7},
  pages   = {073140},
  year    = {2025},
  doi     = {10.1063/5.0272051}
}

@article{LeonPazo2022PRE,
  author  = {Iv{\'a}n Le{\'o}n and Diego Paz{\'o}},
  title   = {Enlarged {Kuramoto} model: Secondary instability and transition to collective chaos},
  journal = {Phys. Rev. E},
  volume  = {105},
  number  = {4},
  pages   = {L042201},
  year    = {2022},
  doi     = {10.1103/PhysRevE.105.L042201}
}

@article{HongMartens2022Chaos,
  author  = {Hyunsuk Hong and Erik A. Martens},
  title   = {First-order like phase transition induced by quenched coupling disorder},
  journal = {Chaos},
  volume  = {32},
  number  = {6},
  pages   = {063125},
  year    = {2022},
  doi     = {10.1063/5.0078431}
}

@article{BickPanaggioMartens2018,
  author = {Christian Bick and Mark J. Panaggio and Erik A. Martens},
  title   = {Chaos in {Kuramoto} oscillator networks},
  journal = {Chaos},
  volume  = {28},
  number  = {7},
  pages   = {071102},
  year    = {2018},
  doi     = {10.1063/1.5041444}
}

@article{OlmiPolitiTorcini2010,
  author  = {Simona Olmi and Antonio Politi and Alessandro Torcini},
  title   = {Collective chaos in pulse-coupled neural networks},
  journal = {Europhys. Lett.},
  volume  = {92},
  number  = {6},
  pages   = {60007},
  year    = {2010},
  doi     = {10.1209/0295-5075/92/60007}
}

@article{BuzsakiWang2012,
  author  = {Gy{\"o}rgy Buzs{\'a}ki and Xiao-Jing Wang},
  title   = {Mechanisms of gamma oscillations},
  journal = {Annu. Rev. Neurosci.},
  volume  = {35},
  pages   = {203--225},
  year    = {2012},
  doi     = {10.1146/annurev-neuro-062111-150444}
}

@article{Wiesenfeld1998,
  author  = {Kurt Wiesenfeld and Pere Colet and Steven H. Strogatz},
  title   = {Frequency locking in {Josephson} arrays: Connection with the {Kuramoto} model},
  journal = {Phys. Rev. E},
  volume  = {57},
  number  = {2},
  pages   = {1563--1569},
  year    = {1998},
  doi     = {10.1103/PhysRevE.57.1563}
}

@article{SiebererWachtelAltmanDiehl2016,
  author  = {Lukas M. Sieberer and Gil Wachtel and Ehud Altman and Sebastian Diehl},
  title   = {Lattice duality for the compact Kardar--Parisi--Zhang equation},
  journal = {Physical Review B},
  volume  = {94},
  number  = {10},
  pages   = {104521},
  year    = {2016},
  doi     = {10.1103/PhysRevB.94.104521}
}

@article{BrownMoehlisHolmes2004,
  author  = {Brown, Eric and Moehlis, Jeff and Holmes, Philip},
  title   = {On the Phase Reduction and Response Dynamics of Neural Oscillator Populations},
  journal = {Neural Computation},
  year    = {2004},
  volume  = {16},
  number  = {4},
  pages   = {673--715},
  month   = apr,
  issn    = {0899-7667},
  doi     = {10.1162/089976604322860668},
  url     = {https://direct.mit.edu/neco/article/16/4/673-715/6826}
}

@book{ErmentroutTerman2010,
  author    = {Ermentrout, G. Bard and Terman, David H.},
  title     = {Mathematical Foundations of Neuroscience},
  year      = {2010},
  publisher = {Springer},
  address   = {New York},
  series    = {Interdisciplinary Applied Mathematics},
  volume    = {35},
  isbn      = {978-0-387-87708-2},
  doi       = {10.1007/978-0-387-87708-2},
  url       = {https://doi.org/10.1007/978-0-387-87708-2}
}

@article{Winfree1967,
  author  = {Winfree, Arthur T.},
  title   = {Biological Rhythms and the Behavior of Populations of Coupled Oscillators},
  journal = {Journal of Theoretical Biology},
  year    = {1967},
  volume  = {16},
  number  = {1},
  pages   = {15--42},
  month   = jul,
  issn    = {0022-5193},
  doi     = {10.1016/0022-5193(67)90051-3},
  url     = {https://www.sciencedirect.com/science/article/pii/0022519367900513}
}

@incollection{NeteaCMNS2012,
  author    = {Netoff, Theoden I. and Schwemmer, Michael A. and Lewis, Timothy J.},
  title     = {Experimentally Estimating Phase Response Curves of Neurons: Theoretical and Practical Issues},
  booktitle = {Phase Response Curves in Neuroscience: Theory, Experiment, and Analysis},
  editor    = {Schultheiss, Nathan W. and Prinz, Astrid A. and Butera, Robert J.},
  series    = {Springer Series in Computational Neuroscience},
  volume    = {6},
  pages     = {95--129},
  publisher = {Springer},
  address   = {New York, NY},
  year      = {2012},
  doi       = {10.1007/978-1-4614-0739-3_5},
  url       = {https://doi.org/10.1007/978-1-4614-0739-3_5}
}

@article{SchultheissPRE2010,
  author  = {Schultheiss, Nathan W. and Edgerton, Jeremy R. and Jaeger, Dieter},
  title   = {Phase Response Curve Analysis of a Full Morphological Globus Pallidus Neuron Model Reveals Distinct Perisomatic and Dendritic Modes of Synaptic Integration},
  journal = {Journal of Neuroscience},
  year    = {2010},
  volume  = {30},
  number  = {7},
  pages   = {2767--2782},
  month   = feb,
  doi     = {10.1523/JNEUROSCI.3959-09.2010},
  url     = {https://doi.org/10.1523/JNEUROSCI.3959-09.2010}
}

@article{ErmentroutKopell1986,
  author  = {Ermentrout, G. Bard and Kopell, Nancy},
  title   = {Parabolic Bursting in an Excitable System Coupled with a Slow Oscillation},
  journal = {SIAM Journal on Applied Mathematics},
  year    = {1986},
  volume  = {46},
  number  = {2},
  pages   = {233--253},
  doi     = {10.1137/0146017},
  url     = {https://doi.org/10.1137/0146017}
}

@article{Brette2005,
  author  = {Brette, Romain and Gerstner, Wulfram},
  title   = {Adaptive Exponential Integrate-and-Fire Model as an Effective Description of Neuronal Activity},
  journal = {Journal of Neurophysiology},
  year    = {2005},
  volume  = {94},
  number  = {5},
  pages   = {3637--3642},
  doi     = {10.1152/jn.00686.2005},
  url     = {https://doi.org/10.1152/jn.00686.2005}
}

@article{Augustin2017,
  author  = {Augustin, Moritz and Ladenbauer, Josef and Baumann, Fabian and Obermayer, Klaus},
  title   = {Low-Dimensional Spike Rate Models Derived from Networks of Adaptive Integrate-and-Fire Neurons: Comparison and Implementation},
  journal = {PLoS Computational Biology},
  year    = {2017},
  volume  = {13},
  number  = {6},
  pages   = {e1005545},
  doi     = {10.1371/journal.pcbi.1005545},
  url     = {https://doi.org/10.1371/journal.pcbi.1005545}
}

@article{Zerlaut2017,
  author  = {Zerlaut, Yann and Chemla, Sandrine and Chavane, Fr{\'e}d{\'e}ric and Destexhe, Alain},
  title   = {Modeling Mesoscopic Cortical Dynamics Using a Mean-Field Model of Conductance-Based Networks of Adaptive Exponential Integrate-and-Fire Neurons},
  journal = {Journal of Computational Neuroscience},
  year    = {2018},
  volume  = {44},
  number  = {1},
  pages   = {45--61},
  doi     = {10.1007/s10827-017-0668-2},
  url     = {https://doi.org/10.1007/s10827-017-0668-2}
}

@article{diVolo2019,
  author  = {di Volo, Matteo and Romagnoni, Alberto and Capone, Cristiano and Destexhe, Alain},
  title   = {Biologically Realistic Mean-Field Models of Conductance-Based Networks of Spiking Neurons with Adaptation},
  journal = {Neural Computation},
  year    = {2019},
  volume  = {31},
  number  = {4},
  pages   = {653--680},
  doi     = {10.1162/neco_a_01173},
  url     = {https://doi.org/10.1162/neco_a_01173}
}

@article{Carlu2020,
  author  = {Carlu, Mallory and Chehab, Omar and Dalla Porta, Leonardo and Depannemaecker, Damien and H{\'e}ric{\'e}, C{\'e}dric and Jedynak, Mathieu and K{\"o}ksal Ers{\"o}z, Elif and Muratore, Paolo and Souihel, Saoussen and Capone, Cristiano and Zerlaut, Yann and Destexhe, Alain and di Volo, Matteo},
  title   = {A Mean-Field Approach to the Dynamics of Networks of Complex Neurons, from Nonlinear Integrate-and-Fire to Hodgkin--Huxley Models},
  journal = {Journal of Neurophysiology},
  year    = {2020},
  volume  = {123},
  number  = {3},
  pages   = {1042--1051},
  doi     = {10.1152/jn.00399.2019},
  url     = {https://doi.org/10.1152/jn.00399.2019}
}

@article{Luke2013,
  author  = {Luke, Tanushree B. and Barreto, Ernest and So, Paul},
  title   = {Complete Classification of the Macroscopic Behavior of a Heterogeneous Network of Theta Neurons},
  journal = {Neural Computation},
  year    = {2013},
  volume  = {25},
  number  = {12},
  pages   = {3207--3234},
  doi     = {10.1162/NECO_a_00525},
  url     = {https://doi.org/10.1162/NECO_a_00525}
}

@article{Daido1996,
  author  = {Daido, Hiroaki},
  title   = {Multibranch Entrainment and Scaling in Large Populations of Coupled Oscillators},
  journal = {Physical Review Letters},
  year    = {1996},
  volume  = {77},
  number  = {7},
  pages   = {1406--1409},
  doi     = {10.1103/PhysRevLett.77.1406},
  url     = {https://doi.org/10.1103/PhysRevLett.77.1406}
}

@article{Skardal2011,
  author  = {Skardal, Per Sebastian and Ott, Edward and Restrepo, Juan G.},
  title   = {Cluster Synchrony in Systems of Coupled Phase Oscillators with Higher-Order Coupling},
  journal = {Physical Review E},
  year    = {2011},
  volume  = {84},
  number  = {3},
  pages   = {036208},
  doi     = {10.1103/PhysRevE.84.036208},
  url     = {https://doi.org/10.1103/PhysRevE.84.036208}
}

@article{HatchettUezu2008,
  author  = {Hatchett, Jonathan P. L. and Uezu, Tatsuya},
  title   = {Mean field and cavity analysis for coupled oscillator networks},
  journal = {Phys. Rev. E},
  volume  = {78},
  number  = {3},
  pages   = {036106},
  year    = {2008},
  month   = sep,
  doi     = {10.1103/PhysRevE.78.036106}
}

@article{Metz2025SparseDMFT,
  author  = {Metz, Fernando L.},
  title   = {Dynamical Mean-Field Theory of Complex Systems on Sparse Directed Networks},
  journal = {Phys. Rev. Lett.},
  volume  = {134},
  number  = {3},
  pages   = {037401},
  year    = {2025},
  month   = jan,
  doi     = {10.1103/PhysRevLett.134.037401},
  eprint  = {2406.06346},
  archivePrefix = {arXiv},
  primaryClass  = {cond-mat.dis-nn}
}

@software{Reddy_Zenodo_2026,
  author    = {Reddy, Kanishka},
  title     = {Simulation code for ``Compact dynamical mean-field theory of oscillator networks''},
  year      = {2026},
  publisher = {Zenodo},
  doi       = {10.5281/zenodo.18796357}
}

\end{document}